\documentclass[11pt,twoside]{article}
\usepackage{./asp2014}

\def\kms{$\mathrm{km\,s}^{-1}$}

\def\espa{ESPaDOnS}

\def\R{\ensuremath{R_{\odot}}}

\def\bz{$\langle$B$_z\rangle$}
\def\nz{$\langle$N$_z\rangle$}

\aspSuppressVolSlug
\resetcounters

\begin{document}

\title{A comprehensive and critical re-analysis of FORS spectropolarimetric observations: the case of HD\,92207}
\author{L. Fossati,$^1$ S. Bagnulo,$^2$ J.~D. Landstreet,$^3$ and O. Kochukhov$^4$
\affil{$^1$Argelander-Institut f\"ur Astronomie der Universit\"at Bonn, Auf dem Huegel 71, 53121 Bonn, Germany; \email{lfossati@astro.uni-bonn.de}}
\affil{$^2$Armagh Observatory, College Hill, Armagh BT61 9DG, Northern Ireland, UK; \email{sba@arm.ac.uk}}
\affil{$^3$Physics \& Astronomy Department, The University of Western Ontario, London, Ontario N6A 3K7, Canada; \email{jlandstr@uwo.ca}}
\affil{$^4$Department of Physics and Astronomy, Uppsala University, 751 20 Uppsala, Sweden; \email{oleg.kochukhov@physics.uu.se}}}

\paperauthor{L. Fossati}{lfossati@astro.uni-bonn.de}{}{Universit\"at Bonn}{Argelander-Institut f\"ur Astronomie}{Bonn}{}{53121}{Germany}
\paperauthor{S. Bagnulo}{sba@arm.ac.uk}{}{Author2 Institution}{Armagh Observatory}{Armagh}{}{BT61 9DG}{Northern Ireland, UK}
\paperauthor{J.~D. Landstreet}{jlandstr@uwo.ca}{}{The University of Western Ontario}{Physics \& Astronomy Department}{London}{Ontario}{N6A 3K7}{Canada}
\paperauthor{O. Kochukhov}{oleg.kochukhov@physics.uu.se}{}{Uppsala University}{Department of Physics and Astronomy}{Uppsala}{}{751 20}{Sweden}

\begin{abstract}
Stellar spectropolarimetry has become extremely popular during the last decade, and has led to major advances in our understanding of stellar magnetic fields and of their impact on stellar structure and evolution. Many important discoveries have been obtained thanks to observations performed with the FORS low-resolution spectropolarimeters of the ESO Very Large Telescope. We first review and summarise the major results of a homogeneous re-reduction and analysis of all single-slit FORS1 spectropolarimetric observations. This work revealed a non-negligible dependence of the results upon the adopted reduction and analysis procedure, as well as the presence of instabilities, revealing that photon noise is not the only source of uncertainty. As a consequence of our new analysis and assessment of the uncertainties, we are not able to confirm a large number of magnetic field detections presented in the past for a variety of stars. We further summarise the results of FORS2 spectropolarimetric observations of the A0 supergiant HD\,92207 which allowed us to explore further the nature of the instabilities, roughly constraining their maximum impact on the derived Stokes profiles and magnetic field values. We finally present new results obtained with a further independent pipeline on the FORS2 data of HD\,92207, confirming our previous analysis, and discuss simple quality-check controls which can be performed on the data in order to distinguish between genuine and spurious signals. All together, our results reveal that the FORS spectropolarimeters are indeed reliable instruments, when their capabilities are not pushed beyond the limits of a Cassegrain mounted low-resolution spectrograph.
\end{abstract}
\section{Introduction}\label{sec:intro}
A large number of low resolution spectropolarimetric data aiming at detecting stellar magnetic fields have been collected with the FORS (FORS1 and FORS2) spectropolarimeters \citep{appenzeller1998} of the ESO Very Large Telescope (VLT). FORS1 was decommissioned in April 2009 and the polarimetric optics were then moved to FORS2. Both FORS1 and FORS2 are multi-mode optical instruments capable of imaging, polarimetry, and long slit and multi-object spectroscopy. 

Spectropolarimetric observations obtained with both instruments, particularly FORS1, have been used to claim the detection of magnetic fields in several different classes of stars across the Hertzsprung-Russell diagram, including for instance central stars of planetary nebulae \citep{jordan2005}, hot subdwarfs \citep{otoole2005}, $\beta$\,Cephei and slowly pulsating B stars \citep{hubrig2009a}, Be stars \citep{hubrig2009b}, and normal O-type stars \citep{hubrig2008b,hubrig2013a}. Many of these detections have been obtained close to the limit of the 3$\sigma$ level and only some have been confirmed on the basis of further high resolution spectropolarimetric observations.

As a matter of fact, in relation to FORS spectropolarimetric data there are inconsistencies of various nature present in the literature:
\begin{itemize}
\item inconsistencies between field measurements obtained with FORS and other instruments \citep[e.g., \espa; see][]{silvester2009};
\item inconsistencies between the results obtained by different authors from the same FORS dataset \citep[e.g.,][]{mcswain2008,hubrig2009b};
\item inconsistencies between the results obtained by the same author, but at different epochs, from the same FORS dataset \citep[e.g.,][]{wade2005,wade2007a};
\item inconsistencies between field measurements obtained from different subsets within a series of frames \citep[e.g.,][]{hubrig2004a};
\item some global inconsistencies of the full FORS dataset \citep[e.g., high incidence of field detections in the null profile;][]{bagnulo2012}.
\end{itemize}

In Sect.~\ref{sec:firs1} we review the major results and conclusions obtained from a homogeneous re-reduction and analysis of all single-slit FORS1 spectropolarimetric observations, first presented by \cite{bagnulo2012}. In Sect.~\ref{sec:hd92207} we further review the major results obtained from FORS2 spectropolarimetric data of the A0 supergiant HD\,92207, presented by \cite{bagnulo2013}, disproving the detection of a magnetic field, as suggested by \cite{hubrig2012}. We then present the results of a further re-reduction of the FORS2 observations of HD\,92207 using an independent pipeline (Sect.~\ref{sec:mag-field}). In Sect.~\ref{sec:time-var} we speculate upon the possible origin of a spurious Stokes $V$ signal found in the FORS2 observations of HD\,92207. To conclude, we discuss simple quality-check controls that can be performed on the data in order to distinguish between genuine and spurious signals, as well as suggestions aimed at obtaining the highest possible quality single-slit spectropolarimetric observations.

\section{The re-analysis of the ESO FORS1 spectropolarimetric data archive}
\label{sec:firs1}
With the aim of quantifying these inconsistencies and in order to try to identify their origin we performed a re-analyses of all observations gathered in circular polarisation with the FORS1 instrument. We performed the whole analysis using the same technique and tools, namely the ESO FORS pipeline \citep{izzo2010} for the data reduction and a suite of customised {\sc fortran} codes for the measurement of the surface average longitudinal magnetic field (\bz) which was obtained using the following relation \citep[see][]{angel1970,borra1980}:
\begin{equation}
\label{eq:bz}
V(\lambda)=-g_{\rm eff}C_z\lambda^2\frac{1}{I(\lambda)}\frac{{\rm d}I(\lambda)}{{\rm d}\lambda}\langle\\B_z\rangle
\end{equation}
and the least squares technique, first proposed by \citet{bagnulo2002} and further refined by \citet{bagnulo2012}. In Eq.~\ref{eq:bz} $V(\lambda)$ and $I(\lambda)$ are Stokes $V$ and $I$, respectively, $g_{\rm eff}$ is the effective Land\'e factor, which was set to 1.25 except for the region of the hydrogen Balmer lines where $g_{\rm eff}$ was set to 1, and 
\begin{equation}
\label{eq:cz}
C_z=\frac{e}{4 \pi m_ec^2}
\end{equation}
where $e$ is the electron charge, $m_e$ the electron mass, and $c$ the speed of light ($C_z\simeq4.67\,\times\,10^{-13}$\,\AA$^{-1}$G$^{-1}$). \cite{bagnulo2012} also give a discussion of the physical limitations of this technique. 

A typical polarimetric observations is composed of a sequence of spectra obtained alternatively rotating the quarter waveplate from $-$45$^{\circ}$ to $+$45$^{\circ}$ every second exposure (i.e., $-$45$^{\circ}$, $+$45$^{\circ}$, $+$45$^{\circ}$, $-$45$^{\circ}$, $-$45$^{\circ}$, $+$45$^{\circ}$, etc...). All details of the adopted data reduction and analysis of the FORS1 archive are given in \cite{bagnulo2012}. We excluded from this re-analysis all data obtained in ``multi-object'' mode because of problems encountered in the reduction of these data using the ESO FORS pipeline \citep[see][for more details]{bagnulo2012}.

The first results of the re-analysis are given in \cite{bagnulo2012}, while a further work (Bagnulo et al. in prep) will present the magnetic field values we obtained from the whole FORS1 dataset. In \cite{bagnulo2012} we gave a re-assessment of the most controversial magnetic field detections claimed in the past on the basis of FORS1 spectropolarimetric data. In this work we concluded also that the data reduction and analysis procedure has a large impact on both magnetic field and uncertainty values and that the data are occasionally affected by systematics which are not taken into account in the error budget.

To present the effects of adopting different reduction procedures on the finally obtained \bz\ values we showed the results gathered for three stars: a clearly magnetic chemically peculiar A-type star (HD\,94660), a non-magnetic star (HD\,96441), and a star consistently presenting a magnetic field detection around the 3$\sigma$ level (HD\,171184), where for each star we applied twelve slightly different reductions and analysis procedures \citep[see Table~2 of][]{bagnulo2012}. The results we obtained for the latter star (i.e., HD\,171184) are of particular interest: out of the twelve \bz\ values we derived, four show a magnetic field detection at less than 3$\sigma$, seven show a magnetic field detection between 3$\sigma$ and 4$\sigma$, and one shows a magnetic field detection larger than 4$\sigma$. Here we show that 3-to-4$\sigma$ magnetic field detections could be turned into less than 3$\sigma$ detections simply by e.g., trimming the first and last 10 data-points in Stokes $I$ and $V$, or by applying a 2 pixel rebinning \citep[see][for more details]{bagnulo2012}.

In this respect, the major problem is that it is not possible to clearly identify the ``best''/``most appropriate'' data reduction and analysis procedure. For example, one could decide to do a background subtraction or not and it is not clear which of the two options is the correct one, but they would lead to two different \bz\ values, where the difference is not quantifiable {\it a priori}; the situation is similar, for example, for flat fielding. One could also decide to apply a sigma clipping or not, and there are different ways of doing it, all of them equally valid and probably leading to (possibly significantly) different \bz\ values.

In \cite{bagnulo2012} we further performed a statistical analysis of the magnetic field values obtained from the diagnostic null profiles \citep[$N$][]{donati1997,bagnulo2009}, hereafter denoted as \nz, which give and indication of the noise level and are used to highlight the presence/absence of systematic errors in the data \citep[see][for more details]{bagnulo2009}. In Fig.~7 of \cite{bagnulo2012} we presented the histogram of the \bz\ values obtained from the null profiles in the whole FORS1 catalog normalised to their uncertainties. The histogram should take the form of a Gaussian with $\sigma$=1, but we obtained something which resembles a Gaussian with $\sigma$=1.365; this clearly indicates the presence of systematics in the FORS1 data and that photon noise is not the only source of uncertainty. This implicitly explains, at least partly, the number and nature of inconsistencies listed in Sect.~\ref{sec:intro}

These tests and considerations led us to the conclusion that the 3$\sigma$ level cannot be considered as an appropriate detection level for magnetic field measurements obtained with the FORS spectropolarimeters, though this consideration extends in general to all long-slit low resolution spectropolarimeters. On the basis of our large re-analysis and by comparing results obtained with different instruments we concluded that a solid magnetic field detection with the FORS spectropolarimeters can be established only in the presence of {\emph repeated} 5-to-6$\sigma$ detections: as shown below, even a single clear detection ($>$5$\sigma$) might not be sufficient to safely avoid spurious detections.
\section{FORS2 spectropolarimetric observations of HD\,92207}\label{sec:hd92207}
\cite{hubrig2012} presented the results gathered from the analysis of FORS2 spectropolarimetric observations of the A0 supergiant HD\,92207 obtained in three different epochs with the grism 600B. Their analysis showed the detection of a magnetic field at two different epochs: the first one at the $\sim$9$\sigma$ level and the second one at the $\sim$4$\sigma$ level. On the basis of the conclusions given by \cite{bagnulo2012} this detection might appear to be genuine. Nevertheless, the Stokes $V$ profiles shown by \cite{hubrig2012} revealed the presence of several suspiciously strong Stokes $V$ signals at the position of sharp and deep lines (i.e., hydrogen Balmer lines and Ca\,{\sc ii}\,H\&K lines) which appear to be too large for the detected magnetic field strength. For this reason we retrieved the FORS2 data from the ESO archive and re-analysed them in order to confirm/disprove the magnetic field.
\subsection{The spurious detection of a non-existing magnetic field}
\label{sec:mag-field}
We performed the data reduction and analysis using two independent tools. The first one is that used for the re-analysis of the FORS1 archive \citep[see Sect.~\ref{sec:firs1} and][]{bagnulo2012}, while the second one is based on a suite of custom made {\sc iraf}\footnote{Image Reduction and Analysis Facility (IRAF -- {\tt http://iraf.noao.edu/}) is distributed by the National Optical Astronomy Observatory, which is operated by the Association of Universities for Research in Astronomy (AURA) under cooperative agreement with the National Science Foundation.} \citep{tody} and {\sc idl} routines which follow in part the technique and recipes presented by \cite{bagnulo2012,bagnulo2013}.

Here we briefly describe the major characteristics of the latter reduction and analysis procedure/tools while more information will be given in a separate paper (Fossati et al., in prep.). The {\sc iraf} pipeline reduces the raw frames only by removing the bias, i.e. without correcting for the flat-field, and it performs an average extraction of the spectra without background subtraction. Within each dataset, each parallel/perpendicular beam is wavelength calibrated using the parallel/perpendicular beam of one wavelength calibration lamp, usually that obtained with the position angle at -45$^{\circ}$. To make sure that different sets of arc lines are not used for the wavelength calibration, the {\sc iraf} pipeline requests human interaction to calibrate the parallel and perpendicular beams in a completely independent way and it makes sure that always the same arc lines and the same fitting function (i.e., a 3$^{rd}$ order spline) is used for the two beams. After having applied the wavelength calibration, the spectra are binned with the natural sampling of the instrument, 0.75\,\AA/pix in the case of the FORS2 data of HD\,92207.

The extracted and wavelength calibrated spectra are combined by an {\sc idl} routine in order to obtain Stokes $I$, $V$, and the diagnostic $N$ parameter using the ``difference'' method following the formalisms of \citet{bagnulo2009}. One can optionally use the uncertainties given by the {\sc iraf} extraction package ({\sc apall}) or opt for pure photon noise error bars; the latter is the default setting as we noticed that the default {\sc iraf} uncertainties tend to be systematically smaller. The Stokes $V$ profile can be optionally rectified; here we used a 4$^{th}$ order polynomial. A sigma clipping can also be applied on the basis of the $N$ parameter; here we filtered out all data-points where the $N$ profile deviates more than 3$\sigma$ from $\overline{N}$, where $\sigma$ is the standard deviation of $N$. Both \bz\ and \nz\ values are finally calculated by minimising the expression given in Eq.~7 of \cite{bagnulo2012} using either the hydrogen lines, or the metallic lines, or the whole spectrum within the observed wavelength region. The \bz\ and \nz\ uncertainties are then rescaled by the $\chi^2$ as suggested by \cite{bagnulo2012}.

\cite{bagnulo2013} presented the results of the re-analysis of the FORS2 spectra of HD\,92207 that showed the largest \bz\ value in the analysis conducted by \cite{hubrig2012} (i.e., data obtained in 2011). Using the reduction and analysis procedure shown in \cite{bagnulo2012} they obtained a 2.9$\sigma$ magnetic field detection from the Stokes $V$ profile and a 5.7$\sigma$ detection from the $N$ profile. This indicates that a magnetic field signature is not present in the analysed FORS2 data, in contrast to the results of \cite{hubrig2012}. In particular, the clear \nz\ detection indicates the presence of systematics in the data, the origin of which will be discussed in Sect.~\ref{sec:time-var}.

As a further check of this result, we re-analysed the 2011 FORS2 data using the {\sc iraf}/{\sc idl} reduction and analysis tools described above. Figure~\ref{fig:hd-hydrogen} shows the graphical output of the {\sc iraf}/{\sc idl} pipeline obtained when considering the hydrogen lines. The results are given Table~\ref{tab:Bz}. These independent results are in agreement with those obtained by \cite{bagnulo2013} and in clear disagreement with those obtained by \cite{hubrig2012}. For the analysis performed with the {\sc iraf}/{\sc idl} we used the same wavelength range shown by \cite{hubrig2012} and \cite{bagnulo2013}.
\begin{table}[!ht]
\caption{\label{tab:Bz} \bz\ and \nz\ values obtained from the FORS2 data of HD\,92207 using the {\sc iraf}/{\sc idl} pipeline and the while spectrum (2nd and 3rd columns) or the hydrogen lines (4th and 5th columns).}
\smallskip
\begin{center}
\begin{tabular}{lcccc}  
\tableline
\noalign{\smallskip}
MJD & $\langle$B$_z\rangle_\mathrm{all}$ & $\langle$N$_z\rangle_\mathrm{all}$ & $\langle$B$_z\rangle_\mathrm{hyd}$ & $\langle$N$_z\rangle_\mathrm{hyd}$ \\
    & [G] & [G] & [G] & [G] \\
\noalign{\smallskip}
\tableline
\noalign{\smallskip}
55688.168 & $-$249$\pm$83  & $-$281$\pm$78  & $-$182$\pm$99  & $-$254$\pm$93  \\
55936.341 &     25$\pm$129 & $-$210$\pm$119 &     18$\pm$143 & $-$232$\pm$140 \\
56018.224 &  $-$68$\pm$158 &     79$\pm$169 &     95$\pm$182 &      8$\pm$192 \\
\noalign{\smallskip}
\tableline\
\end{tabular}
\end{center}
\end{table}

These results show also that the more ``manual'' {\sc iraf}/{\sc idl} reduction leads to smaller \nz\ detections compared to the analysis computed with the fully automatic ESO FORS pipeline. Such a trend is also present for a few other FORS1 data which showed rather large \nz\ detections in \cite{bagnulo2012} and that we re-analysed with the {\sc iraf}/{\sc idl} pipeline as double-check. We believe that the difference lies in the way the wavelength calibration is performed within the two pipelines. The manual wavelength calibration performed with the {\sc iraf}/{\sc idl} pipeline allows one to make sure the same set of lines is used for both parallel/perpendicular beams, in spite of the fact that this leads to the use of a limited number of calibration lines (usually 6--7 for each beam with the grisms 600B, 600V, and 1200B). On the other hand, the fully automatic ESO FORS pipeline aims instead at maximising the number of lines used for the wavelength calibration, but this often leads to the use of a slightly different set of calibration lines for the two beams, because of their slight difference in transmission/efficiency (i.e., the weakest lines used for the calibration of one beam become too weak to be used for the calibration of the other beam). We are still exploring whether this effect might actually be at the origin of the anomalously large number of \nz\ detections in the FORS1 archive \citep{bagnulo2012}.

We also re-analysed with the {\sc iraf}/{\sc idl} pipeline the FORS2 data of HD\,92207 obtained in 2012, consistently obtaining non-detections, as shown by the results given in Table~\ref{tab:Bz}. We can therefore firmly conclude that there is no evidence of the presence of a magnetic field in the FORS2 data of HD\,92207.
\articlefigure[width=1.0\textwidth]{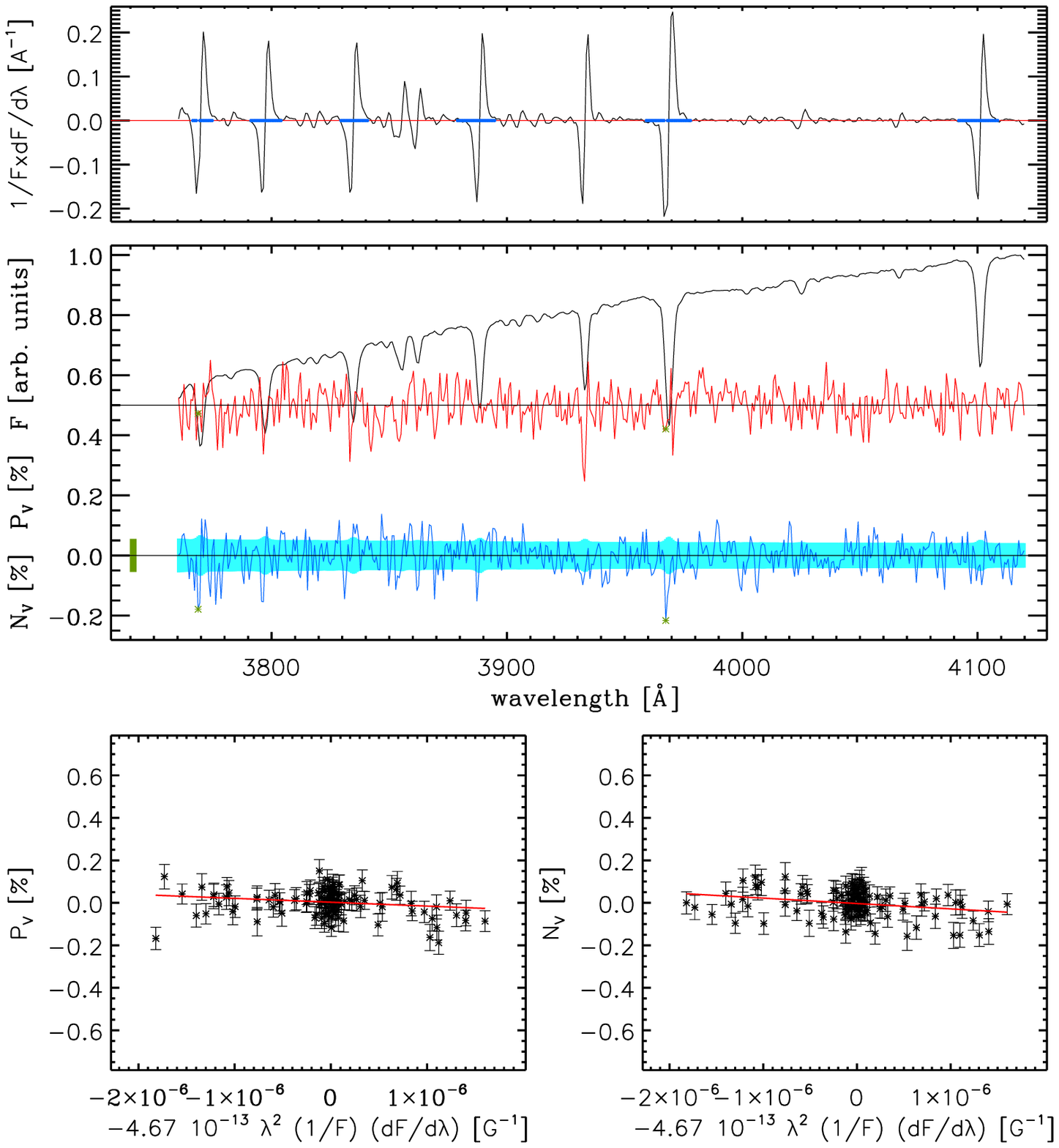}{fig:hd-hydrogen}{Graphical output of the {\sc iraf}/{\sc idl} pipeline obtained from the analysis of the 2011 FORS2 data of HD\,92207 considering the hydrogen lines. Top panel: derivative of Stokes $I$. The regions used for the calculation of the magnetic field are marked by a thick blue line centered at zero. Middle panel: the top profile shows Stokes $I$ arbitrarily normalised to the highest value, the middle red profile shows Stokes $V$ (in \%), while the bottom blue profile shows the $N$ profile (in \%). The green asterisks present on the Stokes $V$ and $N$ spectra mark the points which have been removed by the sigma clipping. The pale blue strip drawn behind the $N$ profile shows the uncertainty associated with each spectral point. As a double-check of the statistical consistency of the data, the thick green bar on the left side of the $N$ spectrum shows the standard deviation of the $N$ profile. Bottom-left panel: linear fit used for the determination of the magnetic field value using Stokes $V$ (i.e., \bz). The red solid line shows the best fit. From the linear fit we obtained \bz\,=\,$-$182$\pm$99\,G. Bottom-right panel: same as the bottom-left panel, but for the null profile (i.e., \nz). From the linear fit we obtained \nz\,=\,$-$254$\pm$93\,G.}
\articlefigure[width=1.0\textwidth]{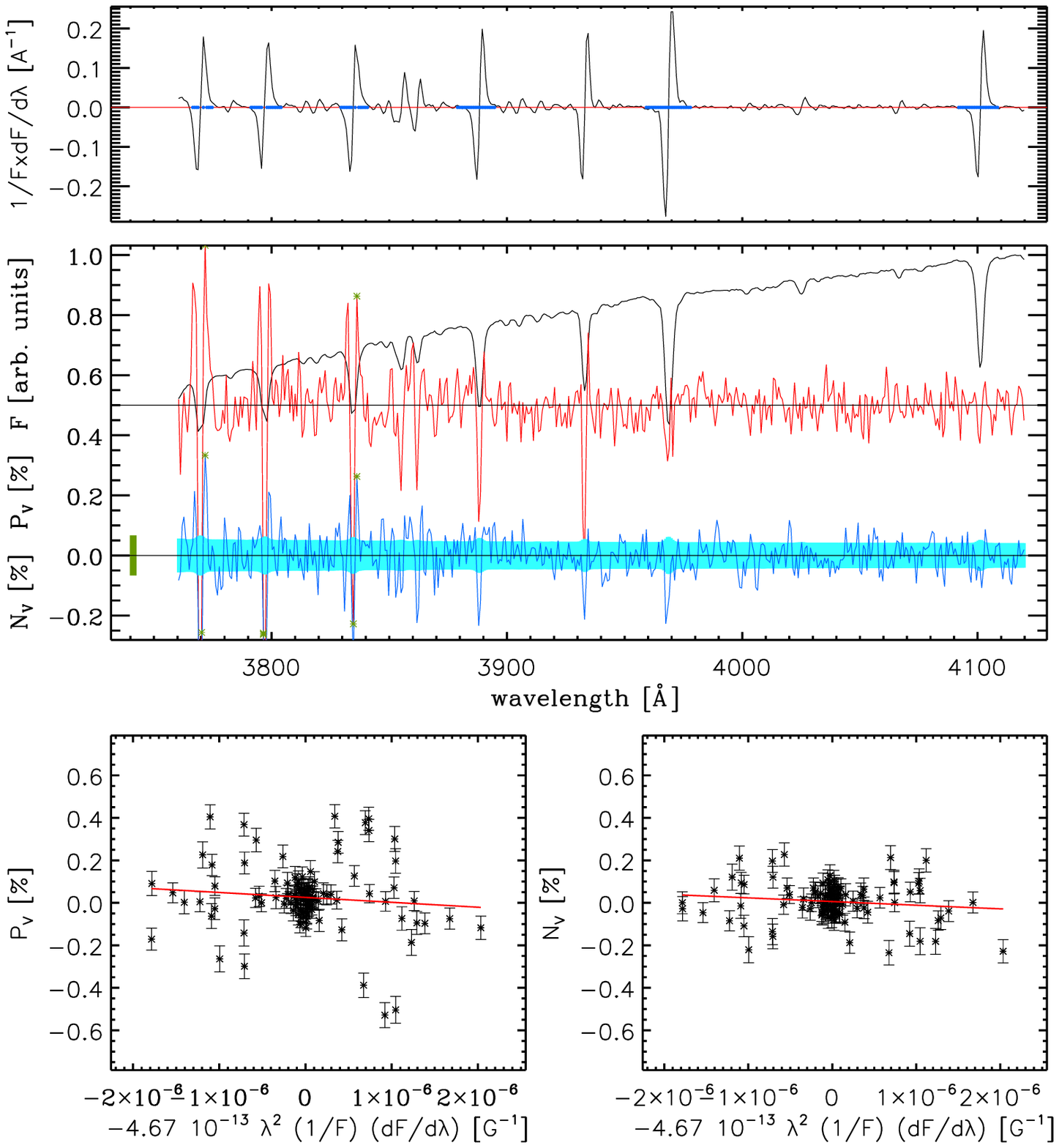}{fig:hd-hydrogen_spikeN}{Same as in Fig.~\ref{fig:hd-hydrogen} but deliberately doing an inaccurate wavelength calibration as described by \cite{bagnulo2013}. From the linear fits we obtained \bz\,=\,$-$242$\pm$211\,G and \nz\,=\,$-$173$\pm$121\,G.}
\articlefigure[width=1.0\textwidth]{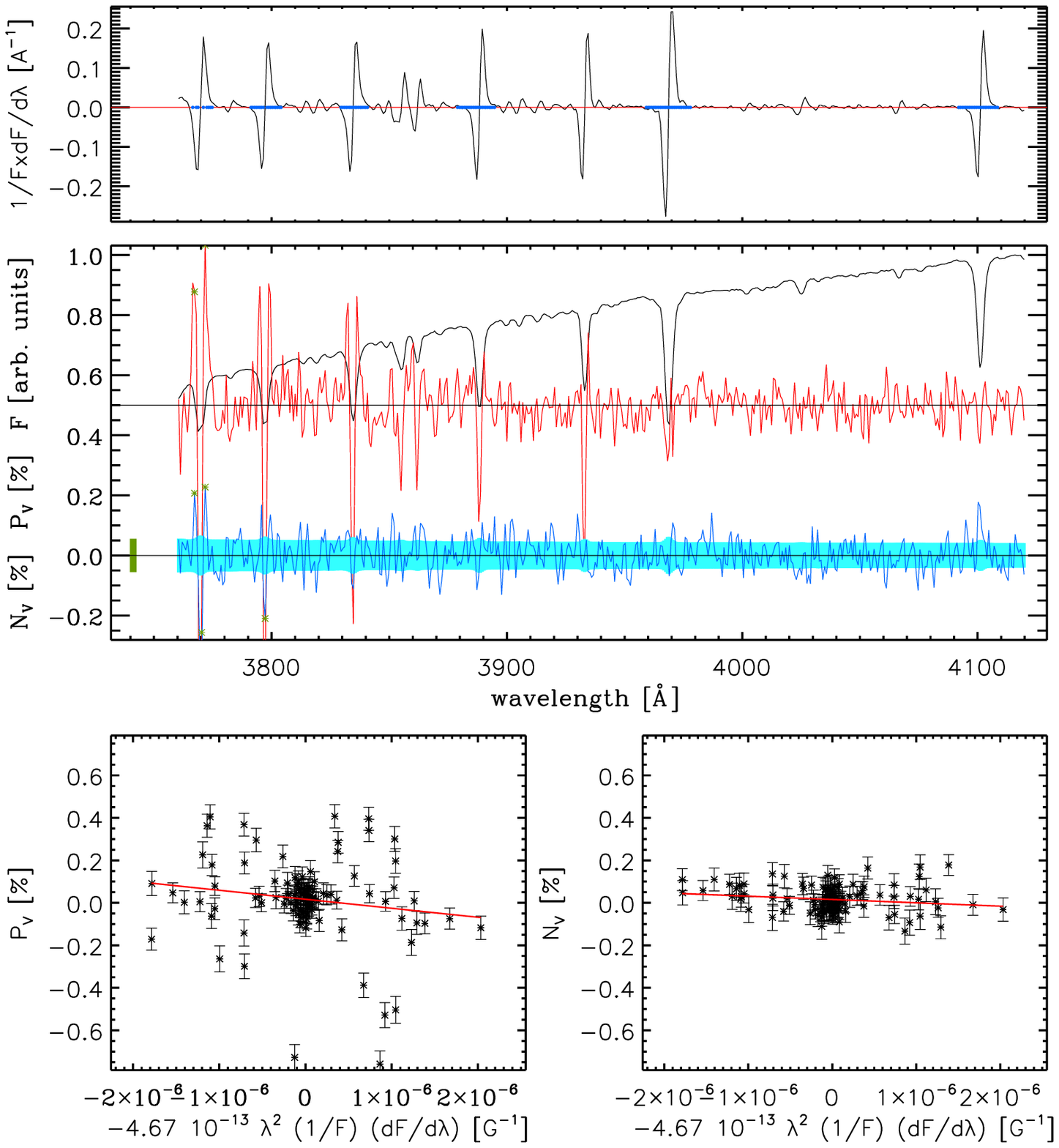}{fig:hd-hydrogen_flatN}{Same as in Fig.~\ref{fig:hd-hydrogen_spikeN} but reshuffling the order of the frames used to derive the Stokes parameters and the $N$ profile as described by \cite{bagnulo2013}. From the linear fits we obtained \bz\,=\,$-$440$\pm$247\,G and \nz\,=\,$-$158$\pm$89\,G. Note that by not rescaling the uncertainties on \bz\ and \nz\ by the $\chi^2$ value we obtained \bz\,=\,$-$440$\pm$74\,G and \nz\,=\,$-$158$\pm$74\,G.}

\cite{bagnulo2013} also presented the results of a magnetic field search for HD\,92207 conducted with the HARPspol spectropolarimeter \citep{snik2011,piskunov2011} feeding the HARPS spectrograph \citep{mayor2003} attached to the ESO 3.6-m telescope in La\,Silla, Chile. The data, obtained in 2013 on four different epochs, were reduced using the {\sc reduce} package \citep{piskunov2002} and analysed using the Least-Squares Deconvolution technique \citep[LSD;][]{donati1997,kochukhov2010}. The analysis of the LSD profiles led to clear non-detections with uncertainties on \bz\ of the order of 10\,G. It is important to notice that, because of the cross-over effect, the presence of a structured magnetic field, particularly of the strength suggested by the results of \cite{hubrig2012}, would be clearly visible in the LSD profiles regardless of the rotation phase. These results firmly indicate that there is no trace of a strong ($>$100\,G) large-scale magnetic field in HD\,92207, in contrast to what was concluded by \cite{hubrig2012}.

Despite our clear results on the non-detection of a magnetic field for this star, it is important to explore the origin of the spurious detection obtained by \cite{hubrig2012}: as a matter of fact, they obtained a $\sim$9$\sigma$ magnetic field detection from the 2011 FORS2 data, well above the ``safety'' threshold of 5--6\,$\sigma$ suggested by \cite{bagnulo2012}. We carefully inspected all major steps performed during the data reduction in order to identify the possible origin of the spurious signal obtained by \cite{hubrig2012}. We were finally able to reproduce \cite{hubrig2012}'s results by deliberately doing an inaccurate wavelength calibration \citep[see][for more details]{bagnulo2013}. It is important to remark that by deliberately doing an inaccurate wavelength calibration we simply introduced a systematic wavelength shift in the parallel beam relative to the perpendicular beam. The Stokes $V$ profiles we obtained in this way \citep[see Fig.~6 of][]{bagnulo2013} present spikes at the position of narrow and deep lines, very similar to those shown by \cite{hubrig2012}. From the analysis of the Stokes $V$ profile we obtained \bz\,=\,$-$325$\pm$105\,G, again very similar to that obtained by \cite{hubrig2012}, except for the uncertainty. Our uncertainty was larger because we rescaled the error bar with the $\chi^2$ of the linear fit. We double-checked the procedure and results described here using the {\sc iraf}/{\sc idl}, obtaining the same profiles and arriving at the same conclusions as in \cite{bagnulo2013}, as shown in Fig.~\ref{fig:hd-hydrogen_spikeN} and Fig.~\ref{fig:hd-hydrogen_flatN}.

Nevertheless, in contrast to what obtained by \cite{hubrig2012}, we obtained similar spikes in the $N$ profile as well, which led to \nz\,=\,$-$355$\pm$75\,G, clearly indicating the spurious nature of the spikes. This result is somewhat reassuring as it shows that the $N$ profile has ``done its job'' by clearly indicating that the Stokes $V$ detection should not be considered to be reliable. Similarly, by rescaling the uncertainty with the $\chi^2$ of the linear fit the detection level obtained from both $V$ and $N$ decreased considerably, showing that the ``safety'' threshold of 5--6\,$\sigma$ suggested by \cite{bagnulo2012} is indeed valid.

We finally attempted to reproduce also the rather flat $N$ profile shown by \cite{hubrig2012}. In the end, by reshuffling the order of the frames used to derive the Stokes parameters and the null profile, we could obtain a much smoother $N$ spectrum, qualitatively similar to that presented by \cite{hubrig2012}. In other words, given a set of spectropolarimetric observations, the $N$ profile is not unique \citep[see][for more details]{bagnulo2013}.
\subsection{On the origin of the short-time line shifts observed in the FORS2 spectra of HD\,92207}
\label{sec:time-var}
In \cite{bagnulo2013} we used the Ca\,{\sc ii}\,K line to show the presence of line shifts, of the order of about a quarter of a pixel ($\sim$4\,\kms), in spectra obtained in consecutive frames. We also noticed that such shifts are consistently present across the whole spectrum, but concentrated on the analysis of the Ca\,{\sc ii}\,K line because of its strength and depth. We considered a range of possible astrophysical and systematic origin of these shifts, but finally excluded all astrophysical (e.g., pulsation) nature of the shifts and attributed them to systematics caused by either instrument flexures or observing conditions (e.g., seeing variations). We showed that these systematic shifts are at the origin of {\it i}) the slight spurious signal which is consistently present in both \bz\ and \nz\ values obtained with both our data reductions and {it ii}) a clear spike in the Stokes $V$ profile at the position of the Ca\,{\sc ii}\,K line \citep[see][for more details]{bagnulo2013}.

\cite{hubrig2014} confirmed what was found by \cite{bagnulo2013} in terms of line shifts, but attributed them to physical changes in the star suggesting short-time pulsations as the most likely cause. To support this conclusion they showed the lack of similar line shifts for two other stars (HD\,93843 and $\zeta$\,Oph) observed in the same nights in which the FORS2 spectra of HD\,92207 were obtained. 

It is here important to recall that each FORS2 observation of HD\,92207 at each position angle was performed with an exposure time of 3 seconds and that between two consecutive exposures in a sequence there is about 1 minute (CCD readout). This implies that, by assuming the line profile variations detected by \cite{bagnulo2013}, and then confirmed by \cite{hubrig2014}, find their origin in the stellar pulsation, the period has to be of the order of a few minutes, at most. Pulsating supergiants, similar to HD\,92207, have typical minimum pulsation periods of the order of days, while for giants, dwarfs, and white dwarfs the typical minimum pulsation period of is of the order to hours, minutes, and seconds, respectively. In other words, it is extremely unlikely, if not even physically impossible, that an A0 supergiant such as HD\,92207 presents pulsations with a period typical of that of main-sequence stars or white dwarfs. In addition, it is really hard to imagine how a star, with a radius of the order of 140\,\R\ \citep{hubrig2012} can show pulsation with a period of the order of minutes; note that a free photon moving on a straight line in the void would need about 5.4 minutes to cover a distance of 140\,\R.

The presence of similar line shifts for all observed lines and the physical characteristics of the star (e.g., radius, rotation rate, temperature, lack of a strong magnetic field) tends to exclude further astrophysical origins, such as surface spots, wind clumping, etc ..., of the observed line shifts. As a result, we have to search for an instrumental or observational systematic origin of the line shifts.

Let us examine the exposure times, adopted slit widths, seeing conditions, and coherence times\footnote{The coherence time is given in the fits header and gives an indication of the timescale of seeing variations.} of the FORS2 observations of HD\,92207, HD\,92843, and $\zeta$\,Oph, the main subject of this work and the two stars used by \cite{hubrig2014} to support their claim that the observed line shifts are of astrophysical origin. Table~\ref{tab:params} lists the relative quantities of each star.
\begin{table}[!ht]
\caption{\label{tab:params} Columns one and three give respectively the slit width (in $\arcsec$) and exposure times (in seconds) adopted for the FORS2 observations of HD\,92207, HD\,92843, and $\zeta$\,Oph. Columns two and four give the sky conditions, namely seeing (in $\arcsec$) and coherence time (in seconds) at the time of the FORS2 observations of HD\,92207, HD\,92843, and $\zeta$\,Oph and listed in the fits headers.}
\smallskip
\begin{center}
\begin{tabular}{lcccc}  
\tableline
\noalign{\smallskip}
Star & Slit  & Seeing & Exposure & Coherence \\
name & width &        & time     & time      \\
     & [$\arcsec$]  & [$\arcsec$]   & [s]      & [s]       \\
\noalign{\smallskip}
\tableline
\noalign{\smallskip}
HD\,92207    & 0.4 & $\sim$0.5 & 3    & 6 \\
\noalign{\smallskip}
\tableline
\noalign{\smallskip}
HD\,92843    & 0.4 & $\sim$0.5 & 30   & 6 \\
$\zeta$\,Oph & 0.4 & $\sim$0.7 & 0.25 & 4 \\
\noalign{\smallskip}
\tableline\
\end{tabular}
\end{center}
\end{table}

For HD\,92207, the FORS2 observations have been performed using a narrow slit and in excellent seeing conditions, namely with a seeing similar to the slit width. In addition, the adopted exposure time was of the same order of the timescale of seeing variations. The most likely explanation of the observed line shifts lies therefore in variations of the position of the star on the slit caused by seeing variations in concurrence with the adoption of a slit width similar to the seeing conditions. From the values listed in Table~\ref{tab:params} it appears then clear why similar line shifts are not observed for both HD\,92843 and $\zeta$\,Oph: for the former the adopted exposure time is much greater than the timescale of seeing variation, while for the latter the adopted exposure time is much shorter than the timescale of seeing variations. For the observations of both HD\,92843 and $\zeta$\,Oph the fact that the adopted slit width is comparable to the seeing does not play a role because of the different timescales between the exposure times and the coherence times. The observed line shifts are most likely also at the origin of the spurious \nz\ detections reported in Sect.~\ref{sec:mag-field}.

\section{Conclusions}\label{conclusions}
We re-reduced and analysed the whole FORS1 ESO archive of long-slit spectropolarimetric observations concluding that a magnetic field detection can be safely reported only on the basis of {\emph repeated} 5--6\,$\sigma$ detections. In addition, we reported the presence of excess noise in the FORS1 data, that we ascribe to systematics of multiple origin (e.g., instrumental, observing conditions, etc ...). We also realised that different, but equally valid, data reduction procedures and tools lead to (significantly) different results \citep[see][for more details]{bagnulo2012}.

We further proceeded to an extensive and deep analysis of the FORS2 observations of HD\,92207 that led \cite{hubrig2012} to report the detection of a rather strong magnetic field (\bz\,$\sim$\,$-$400\,G) at the $\sim$9$\sigma$ level. We re-analysed the FORS2 data using two independent pipelines consistently obtaining non-detections. We managed to reproduce the strong spikes clearly visible in the Stokes $V$ spectra shown by \cite{hubrig2012} only by deliberately doing an inaccurate wavelength calibration that introduced a systematic shift between the parallel and perpendicular beams. Despite this, we obtained a null profile that clearly indicated the spurious nature of the Stokes $V$ spikes. In contrast, \cite{hubrig2012} showed smooth $N$ spectra, that we were able to reproduce by reshuffling the order of the frames used to derive the Stokes parameters and the null profile \citep[see][for more details]{bagnulo2013}. We reported also a non-detection of a magnetic field at the level of a few tens of gauss using high resolution spectropolarimetric HARPSpol data analysed with the LSD technique.

In \cite{bagnulo2013} we also showed the presence of systematic shifts in the line profiles of Stokes $I$ spectra obtained from consecutive frames. The observed line shifts have been further confirmed by \cite{hubrig2014} who attributed them to stellar pulsations. If pulsation is really present and if it causes the observed line profile variations, the pulsation period has to be of the order of a few minutes (5 minutes at most). Such a short pulsation period is typical of dwarfs and white dwarfs and certainly not of supergiants with stellar radii larger than 100\,\R. Further characteristics of the star, such as the lack of a strong magnetic field, the rotation period, and the effective temperature, exclude other astrophysical origin for the line profile variations. In \cite{bagnulo2013} we concluded then that one has to seek their origin in the instrument flexures or in the observing conditions.

As a matter of fact, the observations of HD\,92207 have been obtained with an exposure time comparable to the timescale of seeing variations. This, in addition to the adoption of a slit width similar to the seeing, has led to a non-negligible instability of the star's image on the slit and therefore to the observed line shifts.

We have to remark that the systematics observed for HD\,92207 appear to be of a rather extreme magnitude. Nevertheless, by following our reduction procedures we systematically obtained both a field detection smaller than what recommended by \cite{bagnulo2012} for a reliable detection and a significant signal in the null profile, indicative of the presence of systematics. 

On the basis of our experience we recommend to always re-scale the uncertainties of the \bz\ and \nz\ measurements by the $\chi^2$ value and to report a reliable magnetic field detection only in presence of a {\emph repeated} 5--6\,$\sigma$ detection \citep{bagnulo2012}. We recommend to always look for the absence of line profile shifts in spectra obtained at the same position angle within a single sequence of polarimetric observations similarly to what done by \cite{bagnulo2013}. We also suggest to check the shape of the Stokes $V$ profiles obtained from the single couples of $-$45$^\circ$/$+$45$^\circ$ frames for various lines, paying particular attention to the narrow and deep lines. This will clearly and immediately reveal the presence of systematics which will certainly affect the \bz\ measurements. We finally recommend to always avoid very short exposure times, particularly if the observations are performed with a slit width of comparable size (or larger) than the seeing conditions.

To conclude, our work shows that, assuming one follows a basic set of observational and analysis recommendations, it is indeed possible to confidently use instruments, such as FORS, to measure stellar magnetic fields, even of modest strength.
\acknowledgements 
Luca Fossati would like to deeply thank the staff of the Special Astrophysical Observatory, particularly Iosif Romanyuk, for their warm welcome and hospitality. LF acknowledges support from the Alexander von Humboldt foundation.
\bibliography{fors}  
\end{document}